\pdfoutput=1
\documentclass{JINST}

\title{The distributed Slow Control System of the XENON100 Experiment}

\author{E. Aprile$^a$, M. Alfonsi$^b$, K. Arisaka$^c$, F. Arneodo$^d$, C. Balan$^e$, L. Baudis$^f$, A. Behrens$^f$, P. Beltrame$^c$, K. Bokeloh$^g$, E. Brown$^g$, G. M. Bruno$^d$, R. Budnik$^a$, M. Le Calloch$^h$, \mbox{J. M. Cardoso$^e$\thanks{Corresponding author.},}  W.-T. Chen$^h$, B. Choi$^a$, H. Contreras$^a$, J.-P. Cussonneau$^h$, \mbox{M. P. Decowski$^b$,}  E. Duchovni$^i$, S. Fattori$^j$, A. D. Ferella$^f$, W. Fulgione$^k$, F. Gao$^l$, \mbox{M. Garbini$^m$,} K.-L. Giboni$^a$, L. Goetzke$^a$, C. Grignon$^j$, E. Gross$^i$, W. Hampel$^n$, \mbox{D.N. McKinsey$^q$$^\dag$}, A. Kish$^f$, J. Lamblin$^h$, R. F. Lang$^o$, \mbox{C. Levy$^g$,} K. E. Lim$^a$, \mbox{Q. Lin$^l$,} S. Lindemann$^n$, M. Lindner$^n$, J. A. M. Lopes$^e$, K. Lung$^c$, A. Manzur$^q$$^\dag$, \mbox{T. Marrod\'an Undagoitia$^f$}, F. V. Massoli$^m$, Y. Mei$^p$, \mbox{A. J. Melgarejo Fernandez$^a$,} \mbox{Y. Meng$^c$,} A. Molinario$^k$, E. Nativ$^i$, K. Ni$^l$, U. Oberlack$^{j,p}$, S. E. A. Orrigo$^e$, E. Pantic$^c$, \mbox{J.V. Patricio$^e$}, R. Persiani$^m$, G. Plante$^a$, \mbox{N. Priel$^i$,} A. C. C. Ribeiro$^e$,  \mbox{A. Rizzo$^a$,} S. Rosendahl$^g$, J. M. F. dos Santos$^e$, \mbox{G. Sartorelli$^m$}, \mbox{J. Schreiner$^n$,} \mbox{M. Schumann$^f$,} L. Scotto Lavina$^h$, \mbox{P. R. Scovell$^c$}, M. Selvi$^m$, P. Shagin$^p$, \mbox{H. Simgen$^n$}, \mbox{A. Teymourian$^c$,} D. Thers$^h$, O. Vitells$^i$, H. Wang$^c$, M. Weber$^n$ and \mbox{C. Weinheimer$^g$} \\(The XENON100 Collaboration)\\
\\
\llap{$^a$}Physics Department, Columbia University, New York, NY 10027, USA\\
\llap{$^b$}Nikhef and the University of Amsterdam, Science park, Amsterdam, Netherlands\\
\llap{$^c$}Physics \& Astronomy Department, University of California, Los Angeles, USA\\
\llap{$^d$}INFN, Laboratori Nazionali del Gran Sasso, Assergi, 67100, Italy\\
\llap{$^e$}Department of Physics, University of Coimbra, R. Larga, 3004-516, Coimbra, Portugal\\
\llap{$^f$}Physics Institute, University of Zurich, Winterthurerstr. 190, CH-8057, Switzerland\\
\llap{$^g$}Institut f{\"u}r Kernphysik, Wilhelms-Universit{\"a}t M{\"u}nster, 48149 M{\"u}nster, Germany\\
\llap{$^h$}SUBATECH, Ecole des Mines de Nantes, CNRS/In2p3, Université de Nantes, 44307 Nantes, France\\
\llap{$^i$}Department of Particle Physics and Astrophysics, Weizmann Institute of Science, 76100 Rehovot, Israel\\
\llap{$^j$}Institut f{\"u}r Physik, Johannes Gutenberg Universit{\"a}t Mainz, 55099 Mainz, Germany\\
\llap{$^k$}University of Torino and INFN-Torino, Torino, Italy\\
\llap{$^l$}Department of Physics, Shanghai Jiao Tong University, Shanghai, 200240, China\\
\llap{$^m$}University of Bologna and INFN-Bologna, Bologna, Italy\\
\llap{$^n$}Max-Planck-Institut f{\"u}r Kernphysik, Saupfercheckweg 1, 69117 Heidelberg, Germany\\
\llap{$^o$}Department of Physics, Purdue University, West Lafayette, IN 47907, USA\\
\llap{$^p$}Department of Physics and Astronomy, Rice University, Houston, TX 77005 - 1892, USA\\
\llap{$^q$}Department of Physics, Yale University, P.O. Box 208120, New Haven, CT 06520, USA\thanks{non-collaboration authors, participation in XENON10 collaboration}\\
 E-mail: \email{cardoso@lei.fis.uc.pt}}

\abstract{The XENON100 experiment, in operation at the Laboratori Nazionali del Gran Sasso (LNGS) in Italy, was designed to search for evidence of dark matter interactions inside a volume of liquid xenon using a dual-phase time projection chamber. This paper describes the Slow Control System (SCS) of the experiment with emphasis on the distributed architecture as well as on its modular and expandable nature. The system software was designed according to the rules of Object-Oriented Programming and coded in Java, thus promoting code reusability and maximum flexibility during commissioning of the experiment. The SCS has been continuously monitoring the XENON100 detector since mid 2008, remotely recording hundreds of parameters on a few dozen instruments in real time, and setting emergency alarms for the most important variables.}


\begin{document}

\section{Introduction}

XENON100 is a three-dimensional position-sensitive dual-phase (liquid/gas) time projection chamber (TPC) \cite{Aprile_2012a} filled with ultra-pure liquid xenon to perform a low background dark matter search. The detector is installed underground at Laboratori Nazionali del Gran Sasso (LNGS) in Italy. Particle interactions in the sensitive volume are detected by two arrays of photomultiplier tubes which detect both primary scintillation and ionization signals. These signals are used to perform position reconstruction in the TPC, to define a fiducial volume allowing for a drastic reduction of the background rate. Nuclear recoils of particles such as neutrons or dark matter WIMPs (weakly interacting massive particle) have a higher ionization density than electronic recoils of $\gamma$ and $\beta$ backgrounds, an effect which can be used to discriminate signal against background \cite{Aprile_2012a}\cite{Aprile_2010}\cite{Aprile_2011}. Results from 225 live days of data of XENON100 show no evidence for dark matter and new limits are imposed for weakly WIMP nucleon scattering cross sections \cite{Aprile_2012b}. These results make XENON100 one of the best detectors currently operational for direct dark matter search.

As particle physics detectors such as this have become increasingly more complex and demanding, special attention has been given to monitor the dozens of system variables in order to guarantee detector integrity and stability. These systems are generically termed "Slow Control Systems" (SCS) and are responsible for the constant verification of variables such as pressure, temperature, high voltages, flow or concentrations, among others, at sampling rates below the Hz range. The main purpose of a SCS in a physics experiment can be divided in two main tasks; the monitoring and control of the conditions potentially able to influence and cause deviations on physics data signals, and the safety management of the most delicate parts of the experimental setup. In both cases, a prompt control operation is required if SCS data deviate from the predefined ranges, either with an automatic procedure or through an operator driven action. This is a particular sensitive aspect for XENON100 since the detector is operated at pressures of $\approx$2.23 atm and at cryogenic conditions (around 181.5 K). Temperature and pressure stability are crucial for the operation of a double phase TPC. This most critical aspect of detector operation is temperature regulated by a PID-controller (Lakeshore 340 \cite{Lakeshore}) with UPS backup \cite{Aprile_2012a}. This closed-loop control is monitored by the SCS through a serial RS232 interface. Indeed, control loops of SCS are kept within the instruments level (i.e. temperature controller or flow controller in the purification circuit) and not directly by the server itself. Finally, one should emphasize the important role of the SCS on the continuous data trend analysis as a proactive tool towards a regular detector operation \cite{Schafer}.     

There were two main constraints involved in the design and construction of the SCS that are related to the diversity of the multi-institutional XENON100 collaboration. The first one refers to the multitude of equipment, and its different interfaces, brought in by the participant institutions. The interface problem was reduced by the use of standard communication protocols available on the instruments. The second constraint is related to the heterogeneity of the software operating systems of the research teams. This is bypassed using cross-platform frameworks, like the Java Virtual Machine environment.  Adding more instruments, with standard interfaces to monitor more physical parameters, can be accomplished within the frame of a single server machine, with limited effort and cost on the overall software design. However, the presented system if not tailored to be fully scalable in case the sensors/instruments channels is greatly increased, beyond the single server expansion abilities, and a different approach to the whole system would be required.

The SCS monitors a wide array of detector parameters, such as TPC pressure, temperatures, liquid xenon levels, cryostat insulation vacuum, TPC high voltages, as well as photomultiplier high voltages and currents. It measures gas flow and pressures in the gas purification system and in the gas storage system. In addition, the SCS monitors environmental parameters and auxiliary systems, such as the Radon gas concentration inside and outside the detector's radiation shield, additional temperatures during the detector preparation, the status of the He compressor (used for the pulse tube refrigerator cooling) and the acquisition rates from the DAQ system.

The following sections describe the SCS architecture (section \ref{sec2}) along with the software structure of the complete distributed system (section \ref{sec3}). Experimental parameters read out by the system are presented (section \ref{sec4}) and example results from detector stability studies are briefly presented and discussed (section \ref{sec5}).

\section{Hardware Architecture}
\label{sec2}

The SCS is based on a client-server architecture on a distributed platform with different generic client roles \cite{Gyurjyan_2003}. The server is located at the LNGS underground facilities, near the XENON100 detector, and is responsible for the readout and storage of all data (slow variables) from the instruments and sensors. The need of local operation is only required in an emergency situation, when the operation mode is changed or in case of a maintenance intervention on the sensors and instruments. Run conditions are quite stable and required intervention on the server side is usually minimal. Client programs are distributed worldwide with the purpose of constantly checking for alarm situations and providing a graphical output of the slow variables' progression, as presented in more detail in section \ref{sec3}.

The core of the SCS hardware is the server, which was implemented on a commercial personal computer platform under GENTOO Linux \cite{Gentoo}. The majority of the instruments and sensors used in XENON100 use the serial interface associated to the RS232 protocol. Since the server originally had only one native serial port available, two additional 8-channel RS232 expansion cards (Lava Octopus 550 \cite{Lavalink}) were installed, thus increasing the number of available ports to 17. Since a single instrument can have more than one parameter monitored, the XENON100 set of instruments actually represents 801 different parameters.

\begin{figure}[h]
	\centering
		\includegraphics[width=\textwidth]{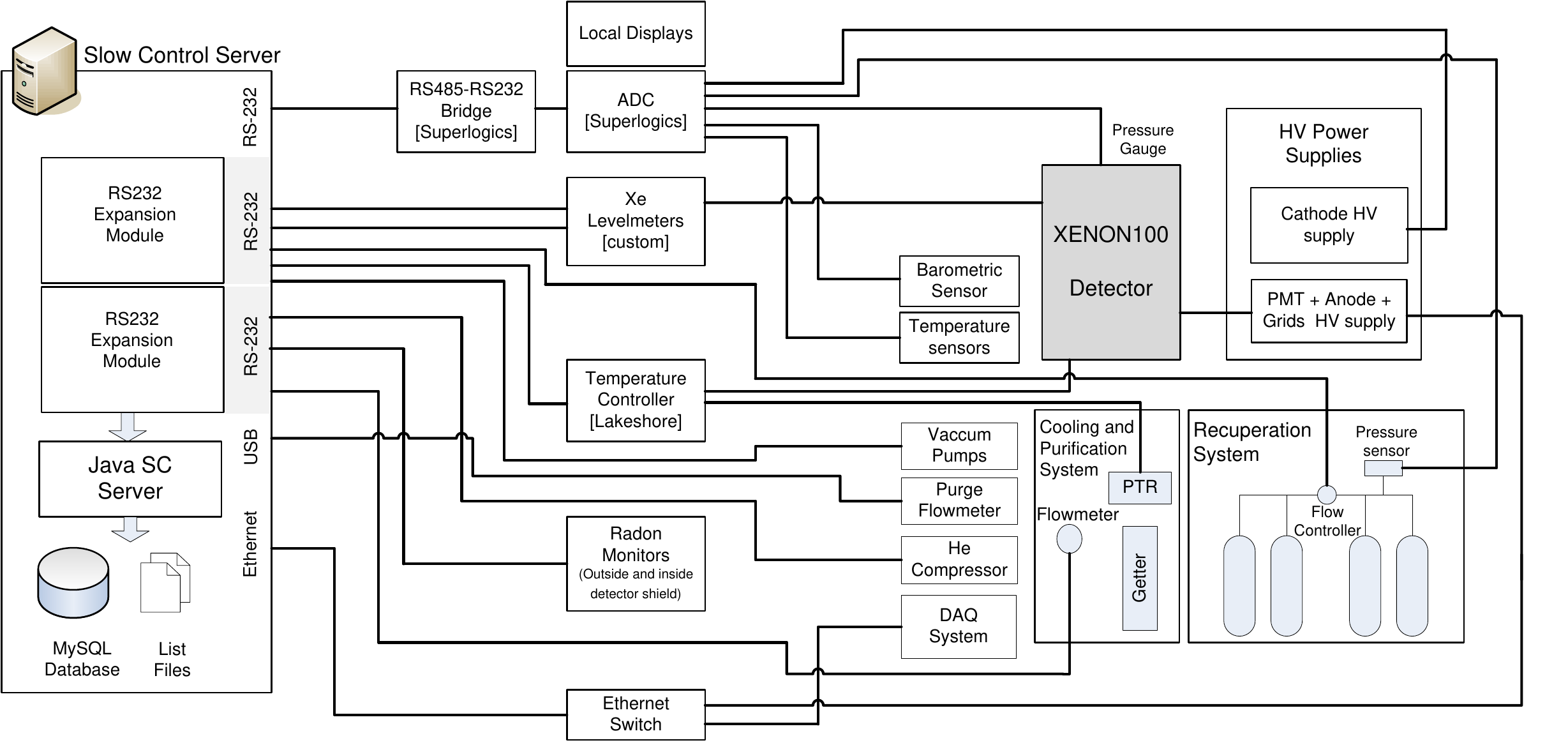}
		\caption{The Slow Control System hardware architecture. Instruments and sensors interface the detector to the SCS server.}
	\label{img:detector_structure}
\end{figure}

In addition to the direct serial interfaces the SCS reads analog signals through an ADC (Superlogics 8017 \cite{Superlogics}) with an RS485-to-RS232 bridge, as well as an N2 flow meter (Voegtlin red-y \cite{redy}) with a USB interface, and an HV supply crate (CAEN SY1527LC \cite{caen}) used for the 242 photomultiplier tubes (PMTs) with an Ethernet interface.  Figure \ref{img:detector_structure} illustrates the physical structure and connections between the SC server and the sensors and instruments.

\section{Software Architecture}
\label{sec3}

The SCS software components are implemented in the object-oriented language Java \cite{Java}. The advantage of using Java consists mostly on the independence from the hardware platform where it is supposed to run. Java ensures that the SCS compatibility is maximized over all the used platforms in the experiment, significantly reducing time and effort on the version management for different operating systems.

\begin{figure}[h]
	\centering
		\includegraphics[width=120mm]{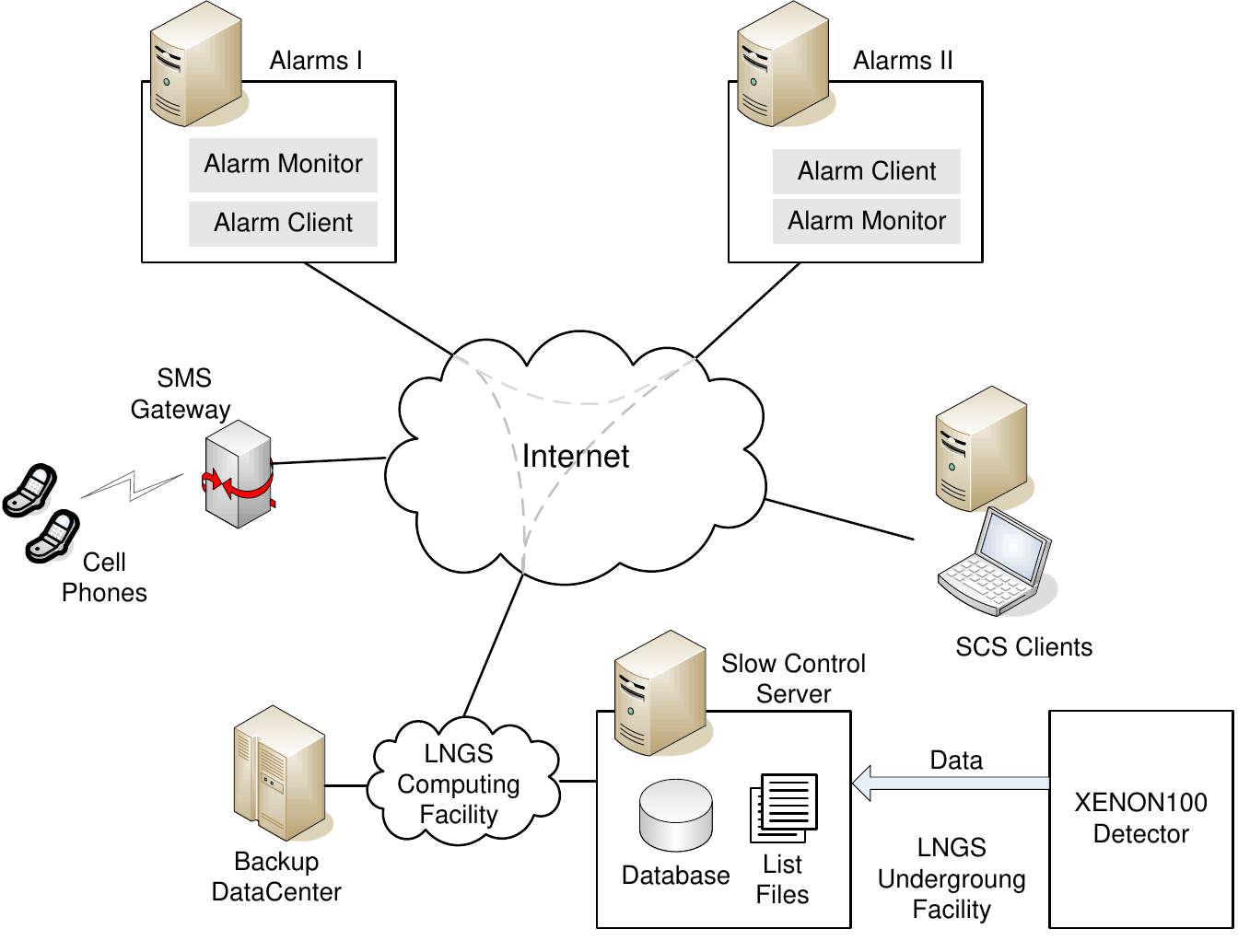}
		\caption{The Slow Control System distributed structure. The SC server is located at the experiment site (LNGS) and SC data is recorded at the server and periodically synchronized at the Backup DataCenter}
	\label{img:sc_system}
\end{figure}

The SCS software structure consists of 4 components. A server runs on a computer close to the detector and locally records the physical parameters making them available to the remote clients. A graphical client program that can run on any machine connected to the internet is used for real time visualization, local recording, and customized local alarms. Two alarm clients continuously monitor the physical parameters read by the SC server. They issue periodic status reports via email and, in case a parameter is outside a pre-defined range, issue alarm messages via email and SMS. An alarm monitor continuously checks the network connection between the alarm clients and the server through a round-trip command. This combination of redundant components guarantees a notification in case of a network problem \cite{Farias_2010}\cite{Zhang_2001}. Figure \ref{img:sc_system} depicts the entire SC distributed system for the XENON100 experiment.

\subsection{Server}

Each instrument read by the SCS can include one or more channels. The measurements for a specific channel are time-stamped and continuously written in log buffers, first in memory and afterwards to a master log in a tree file structure. The instruments are repeatedly polled by the server, as they are triggered by a dedicated timer (instrument specific thread) each with a configurable period. This results in a set of independent timer threads continuously running and logging channel measurements. If one of the instruments fails to respond to a polling request, the overall multithreaded server program is not compromised for the other instruments. 

These measurements are also stored locally in a MySQL database and mirrored to a Backup DataCenter to avoid data loss. This allows direct queries for a particular instrument measurement range during data analysis without introducing latencies due to database queries over the server.

The memory buffered data on the Server can also be accessed by the Client and Alarm Client programs through the Java Remote Method Interface (RMI). The measurement data on the Server is periodically logged to list files and progressively removed from the local memory after a user defined persistence time, to ensure that the memory space is never overloaded.

\subsection{Client}

\begin{figure}[h]
	\centering
		\includegraphics[width=\textwidth]{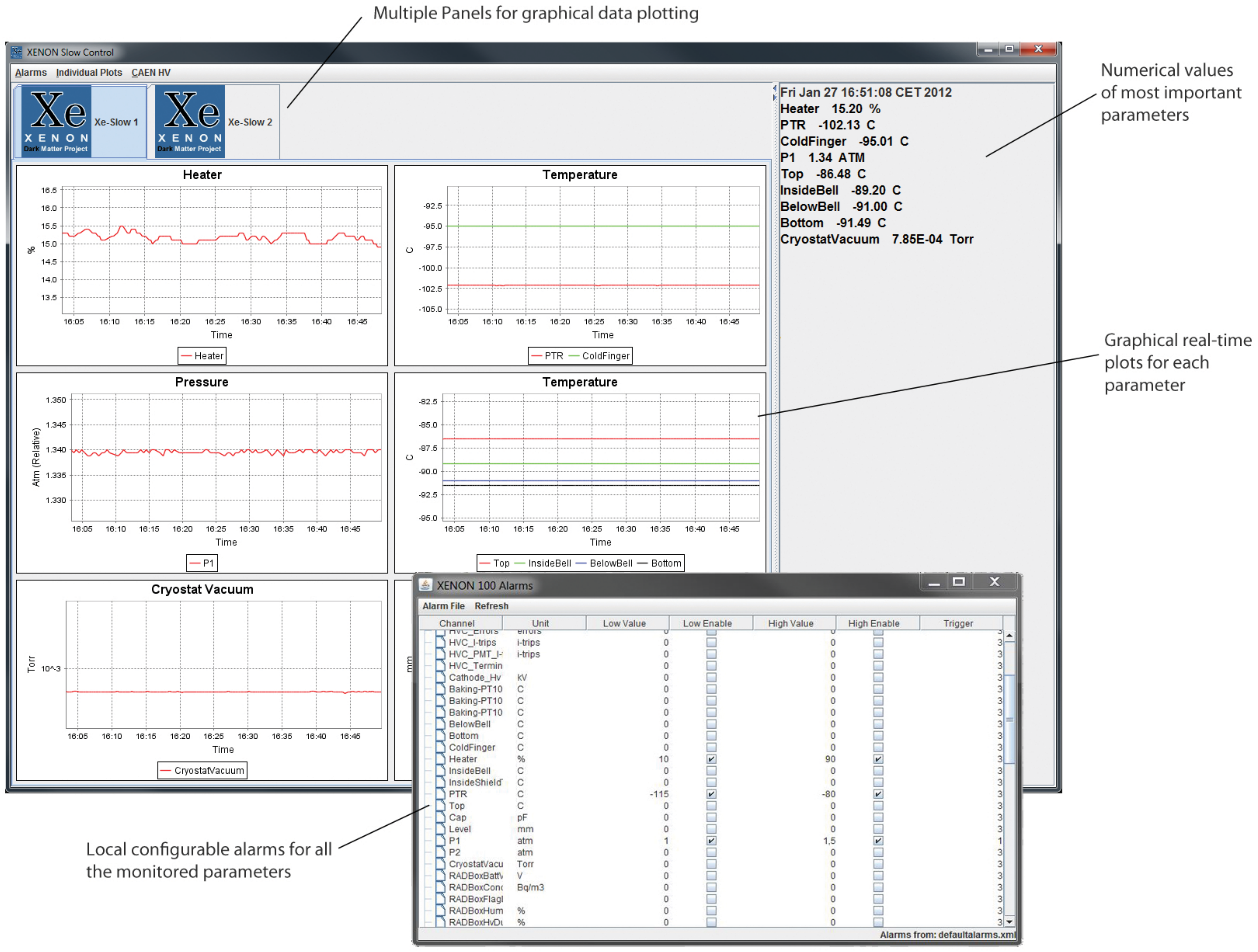}
		\caption{Main window of the SCS client showing some of the most relevant parameters and local alarms settings window.}
	\label{img:client_layout}
\end{figure}

The SC Client application connects to the Server to retrieve the list of channels being monitored and to start timers that will periodically retrieve selected measurements from the Server. Once the list of channels is available, the client is able to register and to display the measurements from each channel. Near real-time channels data is plotted in configurable graphical panels. Independent local alarms can be set on the client but in this case only the local user is warned through a sound message. Figure \ref{img:client_layout} depicts the main window of SC Client.

False alarms due to data glitches are avoided by using a grace count (i.e. the tolerated number of data points outside the allowed range) chosen independently for each parameter.  The Client is also capable of saving data as plots or in text format, allowing to quickly reference data.

Due to the large number of 242 PMTs used in XENON100, plotting is an inefficient way to present the data of these channels. Instead, a table with the set-points and monitored values is given, allowing quick access to the status of a particular PMT.

\subsection{Alarms}

The Alarm Client programs periodically check the values of a pre-defined set of channels chosen by local configuration files. The update period depends on the parameter itself and ranges from a few seconds to around one minute. Alarm events are triggered either if at least one parameter value falls outside the allowed range, if a channel is not updated, or if the connection to the server cannot be established anymore. In either case an alarm state is reached and an email and/or SMS is issued (through an SMS web gateway). Additionally, these programs issue a periodic status report to keep track of the latest values of the most critical parameters for the detector operation such as, e.g., the pressure inside the TPC, temperatures, or the gas purification flow. The alarm configuration (channels to be monitored, emails, phone numbers) are defined locally and alarm clients run independently at two distinct locations to increase redundancy and fault tolerance of the overall system in the case of a network communications failure \cite{Blanke_2003}. The alarms can also be run locally, in which case there's the limitation of absent notification from this alarm client in case of a network failure at LNGS. To avoid such condition, the alarms also cross monitor one another and distribute alarm messages if a network problem or latency is observed (Figure \ref{img:sc_system})

\section{Experimental Parameters}
\label{sec4}

The purpose of the XENON100 Slow Control System is to continuously give a clear picture of the detector status and operation and to allow the user to correlate the science data with the detector operating conditions during data taking.  In the case of XENON100, which employs a liquid xenon TPC \cite{Aprile_2012a}, the key parameters that need to be monitored are those that ensure the detector integrity and the quality of operation, namely:

\begin{itemize}
	\item \textbf{Xenon Gas Pressure} - The pressure of the xenon gas in the system is of crucial importance, both for operation and safety.  The pressure is monitored inside the detector chamber and in the storage system that collects the xenon during detector maintenance.  These are monitored by GT1600 sensors from Stellar Technologies \cite{Stellar_Tech}, digitized by a 16 bit, 8 channel, Superlogics 8017A ADC \cite{Superlogics}. Additionally, there is a physical connection, for the pressure inside the detector, to the underground laboratory system which triggers an alarm for immediate action. This connection also automatically opens a cryogenic solenoid valve to a filled LN$_2$ dewar for emergency cooling of the xenon.
	\item \textbf{Temperatures} - The temperature of the detector is monitored at six locations by calibrated Pt100 sensors. Two of them are located on the Pulse Tube Refrigerator (PTR) providing the power to cool and liquefy the xenon \cite{Haruyama_2006}.  The other four are located at various heights inside the detector (in the liquid xenon and in the gas). All are read by a temperature controller (Lakeshore 340 \cite{Lakeshore}), which communicates its readings to the SC server via an RS232 interface. During detector preparation, an additional set of Pt100 sensors is used to monitor baking temperatures at various external locations on the detector.
	\item \textbf{High Voltage Supply} - The high voltage for the detector is provided by two devices.  A CAEN SY1527LC mainframe supplies the voltages for the anode and all the 242 PMTs, through an Ethernet interface. This device is periodically polled by the SC server that is able to set and read voltages, monitor tripping conditions, raise HV levels without user intervention, store time stamps in the database, and issue alarms if a procedure is not successful after a configurable number of attempts. The HV mainframe interface is made through a set of functions in an HV Wrapper C/C++ library supplied by CAEN \cite{caen}. The cathode is biased by a high precision Heinzinger PNC 100000-3-neg power supply \cite{Heinzinger} and is also read through an ADC channel \cite{Superlogics}.	
	\item \textbf{Flowmeters} - The xenon flow rate is monitored in both the recirculation and recuperation circuits by a mass flow controller (Teledyne Instruments, THPS100) with an RS-232 interface \cite{Teledyne}.
	\item \textbf{Levelmeters} - The xenon level inside the TPC is measured by custom made capacitive sensors both for full-range and fine level assessment at the liquid xenon surface region. Three fine levelmeters and a 4$^{th}$ reference levelmeter, are axially distributed around the TPC axis and, in this way, allow for precise alignment and tilt correction. Readout from these levelmeters is performed through UTI03 chips, from Smartec \cite{Smartec}, which interface the SCS over the RS232 protocol.
	\item \textbf{DAQ Rate} - The SC monitors the average DAQ rate over a local NFS mapped file. The SCS timestamps are correlated to the DAQ system by using the same time server.
\end{itemize}

\vspace{4mm}

Other parameters included in the SCS are summarized in table 1.

\vspace{4mm}

\begin{table}[h]
\centering
\begin{tabular}{l | l l l}
\hline
 \textit{Parameter} & \textit{Instrument} & \textit{Supplier} & \textit{Interface}\\ 
 \hline
 Cryostat Vacuum Pressure \cite{Pfeiffer} & D35614 &	Pfeifer Vacuum &	RS232\\
 Shield LN$_2$ Purge Flow \cite{redy} & Red-y smart series &	Voegtlin Intruments	& Modbus over USB\\
 Radon Concentration \cite{Rad7} (2$\times$) & Rad-7	& Durridge Inc &	RS232\\
 Atmospheric pressure  \cite{sensortechnics} & 144SU005A	& SensorTechnics	& Analog to ADC\\
 Helium Compressor \cite{cryomech} & PT407-CP2880 &	Cryomech Inc. &	RS232\\
 \hline
\end{tabular}
\caption{Other instruments and sensors monitored in the XENON100 experiment}
\label{table1}
\end{table}

\section{Results and Future Work}
\label{sec5}

\begin{figure}[h]
	\centering
		\includegraphics[width=0.7\textwidth]{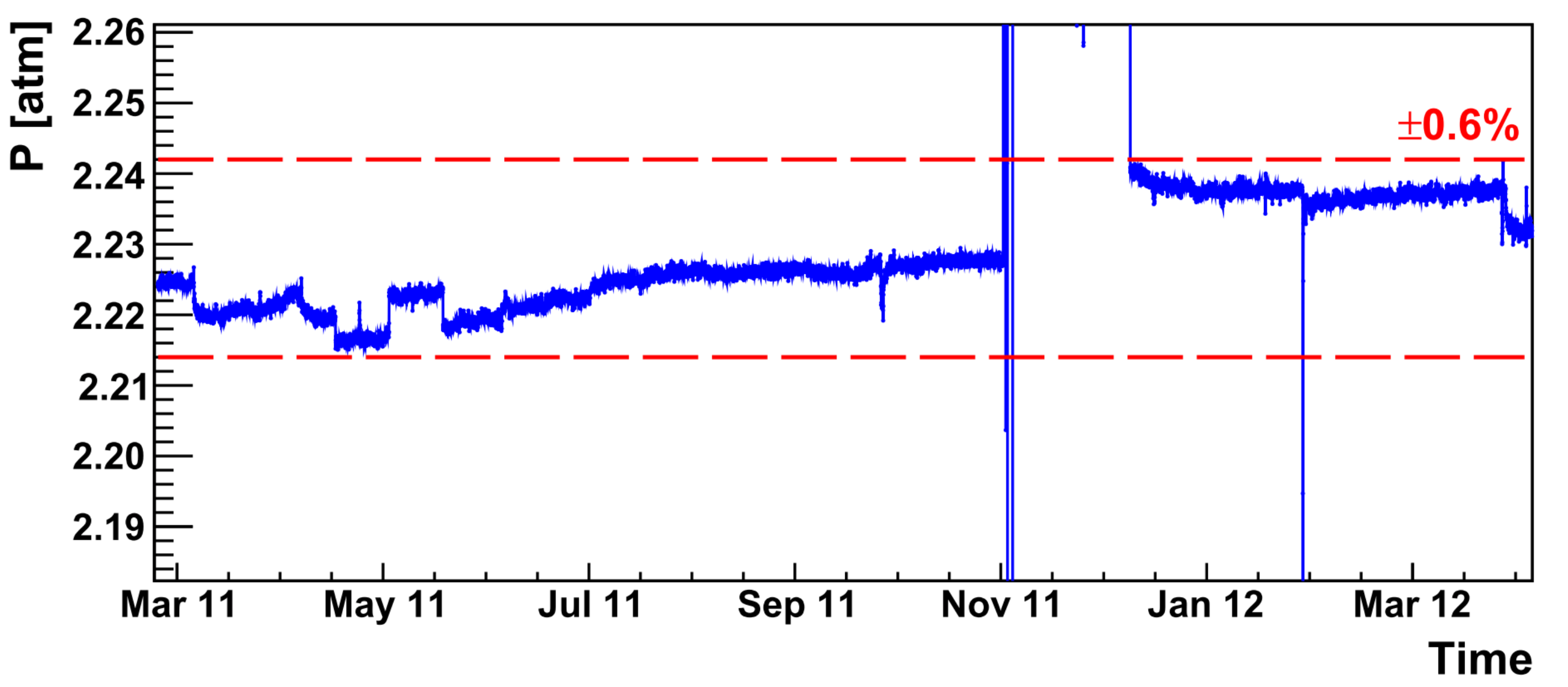}
		\includegraphics[width=0.7\textwidth]{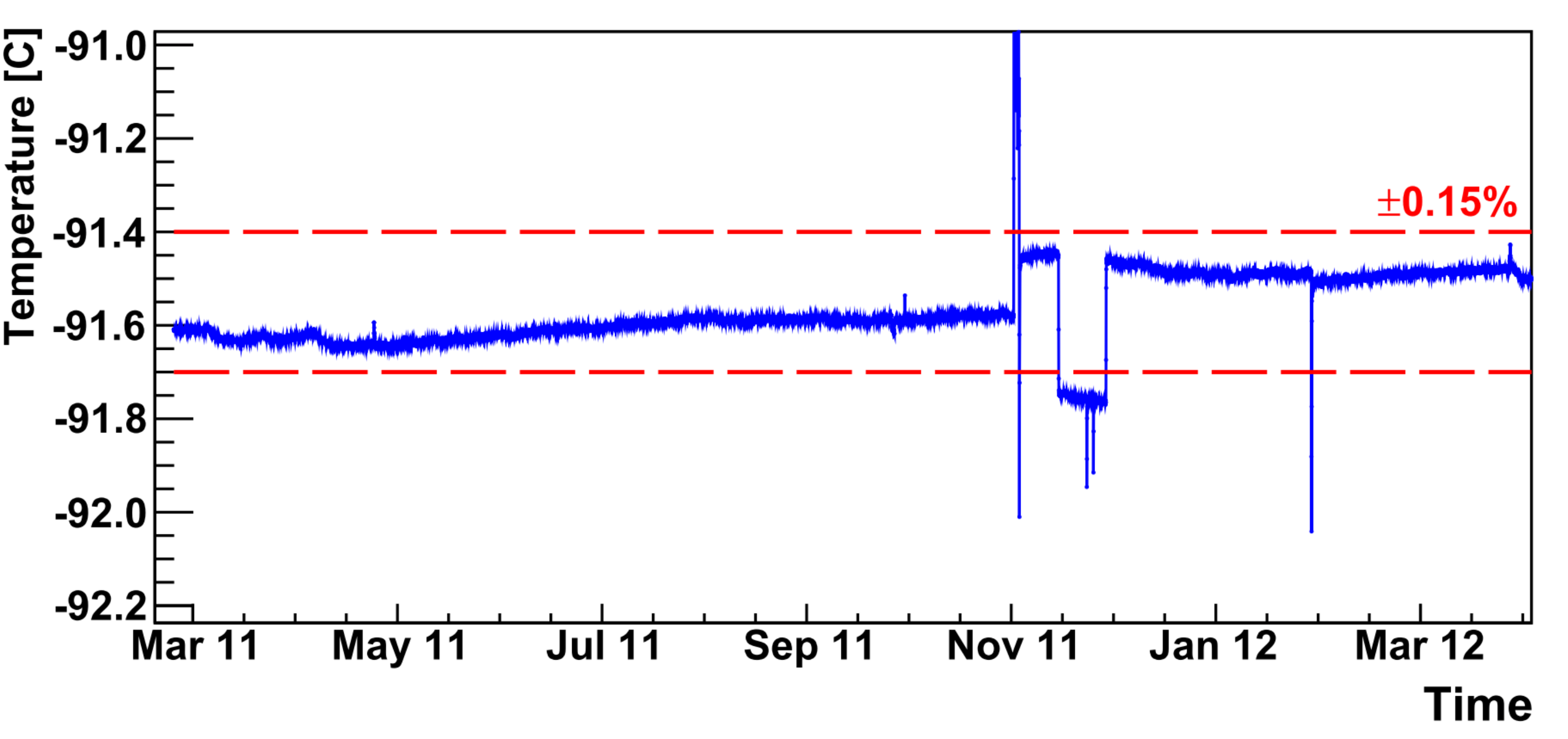}
	\caption{Absolute pressure (top) and temperature (bottom) inside the XENON100 TPC acquired during 13 months of continuous data taking for the results presented in \cite{Aprile_2012b}. The parameters are stable well below $\pm0.6\%$ and $\pm0.15\%$ for the absolute pressure and the temperature, respectively, as indicated by the horizontal lines. All sudden changes of the numbers are related to instrument maintenance and these periods are not used for
data analysis. When smaller time scales are considered, the detector stability is even more remarkable}
	\label{img:run10}
\end{figure}

Data provided by the SCS is promptly accessible through the database or the list files. It is analyzed and cross-correlated with other experimental data to emphasize a particular aspect of the detector operation like the search for trends or deviations from set points. Shifters on-site and scientists abroad can easily follow the LNGS operations online without latency. As an example, the plots for the absolute xenon gas pressure and temperature of the LXe inside the TPC are depicted in Figure \ref{img:run10}, evidencing a remarkable stability (better than $\pm0.6\%$ for the absolute pressure and $\pm0.15\%$ for the temperature) for nearly 13 months of data taking that resulted in the dark matter results presented in \cite{Aprile_2012b}.

Global short term trends can be spotted by a quick observation of this kind of data. Long term stability is also easily monitored by querying the database over the desired time period. Besides maintenance and upgrades, this system has been continuously and steadily running since the experiment's commissioning in early 2008.
Future work tasks for the SCS of XENON100 include the integration of additional instrumentation used on the krypton removal column (used to achieve ultra pure xenon), the improvement of the configuration management, and the creation of a higher control layer able to show the status of each component of the SCS in real-time.

The experience resulting from the development, operation, and maintenance of the XENON100 SCS enters ongoing development of a slow control system for the next experiment XENON1T.

\section{Conclusions}

The Slow Control System (SCS) for the XENON100 direct dark matter search experiment was presented and discussed. The system instrumentation is based on industrial and customized hardware while the software layer was designed using the object-oriented paradigm and coded in Java for platform independence. A timer-thread driven approach enables the readout of a large number of instruments without compromising the overall performance and increases the fault tolerance in case of an instrument failure. The distributed architecture greatly improves redundancy and allows remote monitoring of the detector. The combination of these two features allows for continuous monitoring and safe operation of the XENON100 detector.

\acknowledgments

We acknowledge Richard Hasty and Yamashita Masaki for their contribution towards defining the architecture of this Slow Control System. We also gratefully acknowledge support from NSF, DOE, SNF, Volkswagen Foundation, FCT (project fund PTDC/FIS/100474/2008), Région des Pays de la Loire, STCSM, DFG, Minerva Gesellschaft, and GIF. We are grateful to LNGS for hosting and supporting XENON.

\end{document}